# Possibility of Gravitational Tempering in Colloidal Epitaxy to Obtain a Perfect Crystal

Atsushi Mori,*[1] Yoshihisa Suzuki,[1] and Shigeki Matsuo[1]

[1]*Institute of Technology and Science, 2-1 Minamijosanjima-cho, Tokushima 770-8506*

We have performed Monte Carlo simulations of hard spheres on a pattern under gravity. We have found that a crystal formed at a moderate gravity strength contains essentially no defects while one formed at a higher strength gravity contains a significant amount of defects. This result suggests the possibility of using gravitational-tempering in a colloidal epitaxy to reduce the number of defects in colloidal crystals. Moreover, we wish to emphasize the possibility to obtain a perfect crystal.

Colloidal epitaxy was first proposed in 1997 by van Blaaderen *et al.*[1] as a method to reduce stacking disorders in colloidal crystals. The basic idea of colloidal epitaxy is that the uniqueness of the stacking sequence for face centered cubic (fcc) (001) stacking reduces stacking disorders. The fcc (001) stacking is forced in the colloidal epitaxy by a pattern on the substrate. Additionally, Zhu *et al.*[2] in 1997 found that gravity reduces the stacking disorder in hard-sphere (HS) colloidal crystals. We have already found a glide mechanism for shrinking a stacking fault through a Shockley partial dislocation in fcc (001) stacking in HS crystals by Monte Carlo (MC) simulations.[3] This mechanism has also been suggested by MC simulations of a colloidal epitaxy on a square pattern under gravity.[4,5] In the present work, we note the possibility of gravity-induced tempering based on MC simulations. The number of defects is reduced with increasing gravity in a rage of small gravity. But, it is reduced with decreasing gravity for a certain range. Namely, the relationship between the number of defect and the strength of gravity is not monotonic; the amount of defects reaches a minimum at an optimum gravity strength. Up to a certain gravity strength the crystallinity improves with increasing gravity. In particular, no defects have been observed in the bottom region at this gravity. We propose that settling a colloidal crystal at an optimum gravity strength reduces the number of defects. In other words, one can obtain a perfect crystal with the gravitational tempering.

In 1957, the existence of the crystalline phase in an HS system was shown by MC[6] and molecular dynamics (MD)[7] simulations. In the HS system, the phase behavior is governed only by the particle number density. The fluid and the crystalline phases are separated by a coexistence region of $0.494 < \phi < 0.545$[8] (In 1998, the coexistence densities were revised by Davidchack and Laird[9] as $\phi_f = 0.491$ and $\phi_s = 0.542$ by an MD simulation of a direct crystal-fluid coexistence). Here, $\phi \equiv (\pi/6)\sigma^3(N/V)$ is the volume fraction of the HSs with $\sigma$ being the HS diameter, $N$ the number of particles, and $V$ the volume of the system. As the density increases, the HS system crystallizes. Accordingly, the HS colloids, and also the charge stabilized colloids with repulsive screened Coulombic interaction, crystallize by sedimentation.[10] Defects contained in colloidal crystals obtained without invention are, however, inevitable.

In applications such as production of photonic crystals, we need to reduce the amount of defects in the colloidal crystals. The use of a patterned substrate as a template (a method sometimes referred to as a colloidal epitaxy[1]) is promising. Using a template with a square pattern,[11-14] one can force the crystal growth direction to be fcc <001>. In this stacking direction the stacking sequence is unique, so one can largely avoid stacking disorders, which easily occur in fcc <111> stacking. On the other hand, the effect of gravity, which reduces the number of stacking disorders in colloidal crystals,[2] has been studied as well. The crystalline structure formed in a space shuttle was a random hexagonal close pack (rhcp), whereas on the ground, a mixture state of fcc and rhcp was obtained.[2] This suggested that the reduction of stacking disorders was due to a gravitational effect. Though the final structure was not the rhcp/fcc mixture, Kegel and Dhont[15] presented a result supporting this trend.

We have already found a glide mechanism of the Shockley partial dislocation to shrink an intrinsic stacking fault running along the oblique {111} direction in fcc (001) stacking by an MC simulation.[3] One of the shortcomings of this simulation was that the fcc (001) stacking was driven by stress from a small simulation box with the horizontal periodic boundary conditions (PBC). One of the purposes of our successive simulations is to resolve this artifact. We have already successfully performed fcc (001) stacking in MC simulations of colloidal epitaxy on a square pattern.[4,5,16] Namely, the artificial stress (the stress from the artificial PBC simulation box) has been replaced with a more realizable one (that from the pattern on the substrate). One of the other purposes of our simulations was to reproduce the disappearance of the stacking disorders in simulations with a realizable driving force for fcc (001) stacking. We have already observed the vanishing of the separation of the projection of lattice lines.[4,5,16] If a stacking fault runs across a lattice line, the lattice line is split by the Burgers vector of the Shockley partial dislocation. Additionally, we presume that some of these phenomena were not observed due to the smallness of the system in our previous studies. Finding the stacking fault tetrahedra was the one which we have not observed in our previous studies. We have observed triangular defect structures, which have been identified with projections of tetrahedra.[17] The avoidance of polycrystallization[4,5] is another example that was not observed in our previous studies.

The system size was $L_x = L_y = 25.09\sigma$ and $L_z = 1000\sigma$. $N = 26624$ HSs were placed between well-separated top (at $z = L_z$) and bottom (at $z = 0$) walls. The pattern on the bottom wall was the same as that in a previous study.[3] The groove width was $0.707106781\sigma$ and the side to side separation between two adjacent grooves was $0.338\sigma$.[18] The diagonal length of the intersection of the longitudinal and transverse grooves was $0.9999999997\sigma$. Thus, an HS just fit into, but did not fall into, the intersection. While in a previous study[3] we controlled the dimensionless quantity $g^*\equiv mg\sigma/k_BT$ (which is equal to the reduced inverse gravitational length $\sigma/l_g$ with $l_g = k_BT/mg$ and referred to as the gravitational number) step-wise as proposed previously[19] to avoid polycrystallization, in the present simulation, gravity is switched on suddenly. Here, $m$ is the mass of a particle, $g$ the acceleration due to gravity, and $k_BT$ is the temperature multiplied by Boltzmann's constant. Gravity was applied to an initial random configuration and kept constant for $5.12\times10^7$ Monte Carlo cycles (MCCs). Here, one MCC is defined to include $N$ MC particle position moves. We note here that the control of $g^*$ is directly accomplished by changing the rotation velocity with a centrifugation sedimentation method in experiments.[20-23] By definition increasing $g^*$ corresponds to decreasing $T$. Accordingly, increasing $g^*$ can be realized by cooling as well. Thus, the rapid increase of $g^*$ corresponds to quenching, and keeping $g^*$ at a moderate value corresponds to annealing or tempering.

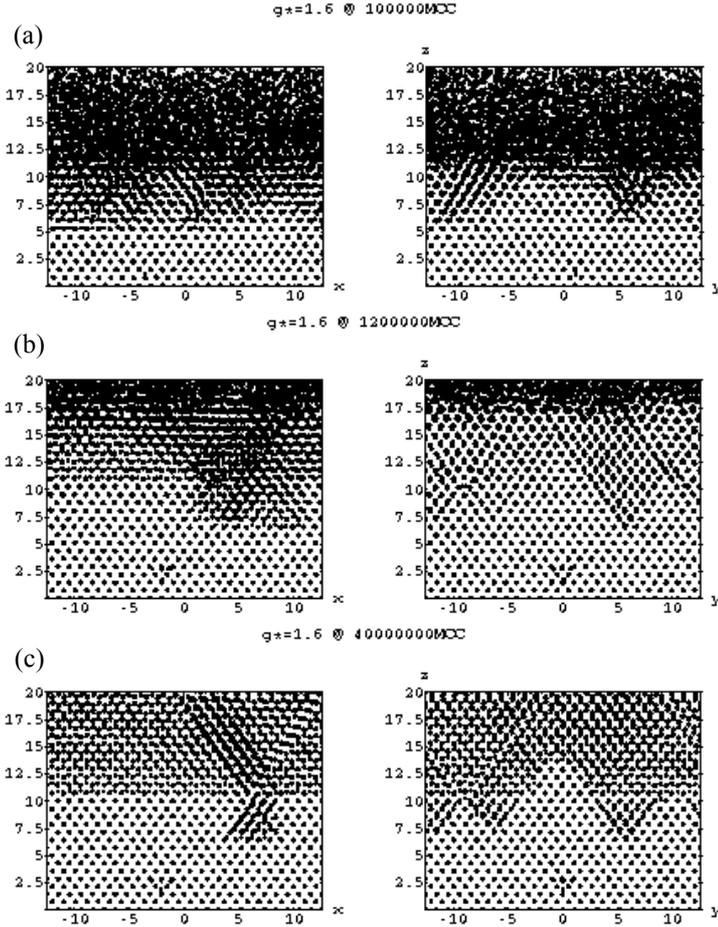

**Figure 1.** Snapshots for $g^*$=1.6: $xz$- and $yz$- projections of particles at (a) $1.0\times10^5$th, (b) $1.2\times10^6$th, and (c) $4.0\times10^7$th MCCs.

We have, so far, performed conventional MC simulations in this series of studies. Transforming them into kinetic MC simulations is simply accomplished by scaling the time.[24,25] That is, MCCs are divided by the acceptance ratio and measured in the unit of $\sigma^2/D_0$, where $D_0$ is the diffusion coefficient in the infinite dilution. This procedure is, however, unnecessary for our present purpose. We have not aimed to evaluate the rate of structural change, such as by looking at the shrinking of the stacking fault or activation barrier.

We have looked at the evolutions of the centers of gravity for $g^*$=0.9-1.6 (previously[5] we looked at for $g^*$=0.6, 0.9, 1.1, and 1.6). We took moving block averages of $z_G \equiv (1/N)\sum_{i=1}^{N} z_i$ for 10000MCCs, where $i$ is the identification number of a particle. In an early stage the centers of gravity $z_G$ sank rapidly. As reported already[5] sedimentation of particles and crystallization were associated with these sinks. After this stage, $z_G$ became nearly constant. However, the $z_G$'s after the rapid dip are slightly different from one another. As observed by following the evolutions of snapshots, defect disappearance took place during those seemingly flat durations (slight decreases in time are seen in magnifications).

We compared snapshots at the end of the simulations at different values of $g^*$. In addition to the avoidance of polycrystallization and the formation of defect structures suggested to be the stacking fault tetrahedra, we found improvement of crystallinity up to $g^*$=1.4 as $g^*$ increases and an increase in the number of defects for $g^*$'s above $g^*$=1.4. As mentioned above, the avoidance of polycrystallization and the formation of the stacking fault tetrahedra were new findings.[5] Looking into the "time" evolutions of snapshots (Figs. 1 and 2), we can confirm these phenomena. The conclusions obtained from Fig. 1 are the same as those in Ref. 5, irrespective of the times at which the snapshots are drawn. In this report, however, we wish to highlight the non-monotonic behavior with respect to the number of defects with increasing $g^*$. Let us compare Figs. 1 (c) and 2 (c). Surprisingly, there are almost no defects in Fig. 2 (c), whereas we see some defects in Fig. 1 (c). We note that Fig. 1 (c) is at $4.0\times10^7$ MCC, which is later than the instance of Fig. 2 (c). It should also be emphasized that, within the simulation duration, the defects which are seen in Fig. 1 (c) were not beginning to disappear. The present results demonstrate a reduction of defects by "heat treatment" at moderate $g^*$. While stacking faults tetrahedra exist in Fig. 2 (b), no stacking fault tetrahedra are observed in Fig. 2 (c). Though the splitting of the projections of lattice lines are observed, any single stacking faults, as observed in Ref. 3, are not observed in Figs. 1 (a) and (b) and Fig. 2 (b). This may be due to the largeness of the system size as more than one stacking disorders overlap in the direction of the projection, resulting in a complicated projection. More than one stacking fault may move and then collide to form a sessile structure, such as a stacking fault tetrahedron. To detect the single stacking fault as has been performed previously,[3] one may limit the observed region before the formation of sessile defect structures.

Now we will discuss our results. We have already discussed that the avoidance of polycrystallization may be the result of a scenario in which a heterogeneous crystalline nucleation on the pattern occurs in an early stage and successive upward growth overcomes the homogeneous nucleation inside the bulk region.[4,5] The primary event is the vanishing of the stacking fault tetrahedra, which was observed at $g^*$=1.4 (and 1.5). From a crystallographic geometric consideration, the stacking fault tetrahedron is sessile. Therefore, if the crystal including the stacking fault tetrahedron is compressed and results in small inter-particle separations, which prevent enough displacement for a particle lying along the stacking faults to move over a particle underlying the stacking fault, the stacking fault tetrahedron cannot move or vanish. We speculate that this situation occurred at $g^*$=1.6. On the other hand, if $g^*$ is not large enough, the driving force for defect disappearance[26] is too weak to eliminate the defects. Therefore, defects remain for $g^*$ < 1.4. The strength of gravity for $g^*$=1.4 and 1.5 should be moderate. Gravitational aging is possible if we keep the gravitational condition moderate. If we keep the gravitational condition moderate before growing colloidal crystals at a strong gravitational condition, the treatments correspond to gravitational annealing. Treatments such as reducing gravity to a moderate strength after growing the colloidal crystals at a strong gravitational condition correspond to gravitational tempering.

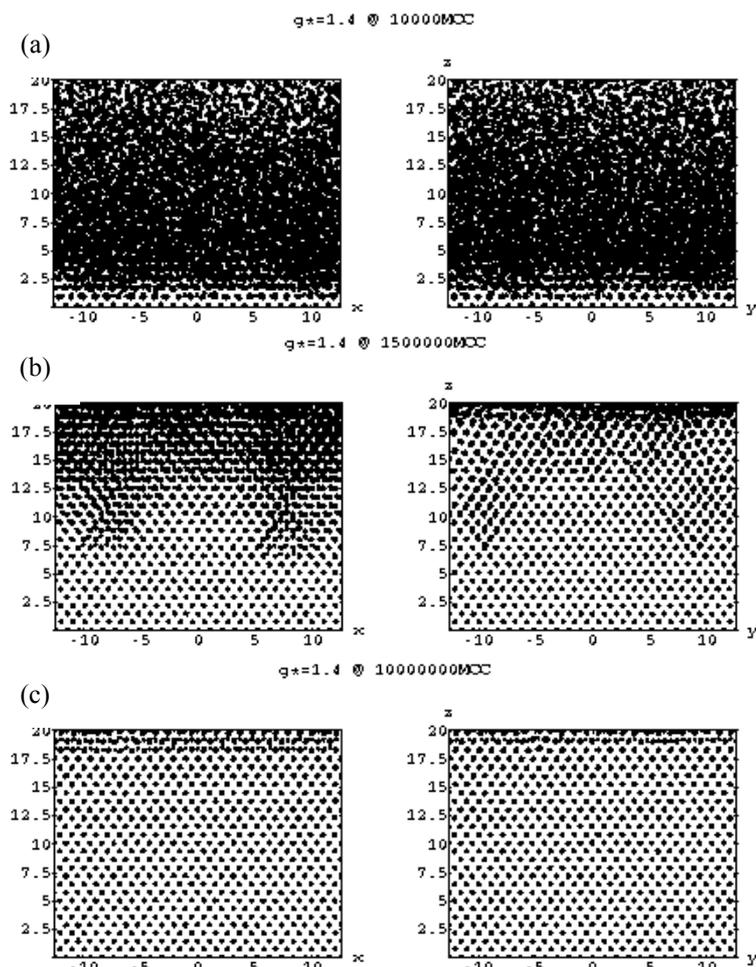

**Figure 2.** Snapshots for $g^*$=1.4: $xz$- and $yz$- projections of particles at (a) $1.0\times10^4$th, (b) $1.5\times10^6$th, and (c) $1.0\times10^7$th MCCs.

Treatments such as "aging" or "annealing," induced by gravity on a flat bottom wall have already been demonstrated. Kegel and Dhont[15] reported that the faulted-twinned fcc structure[27] was obtained by aging for anywhere between two weeks to 40 days in the HS colloidal dispersion of poly(methyl methacrylate) with a $g^*$ on the order of $10^{-2}$-$10^{-4}$. Dolbnya *et al.*[28] found a formation of perfect fcc structure by settling for 1.5-2 years in the HS colloidal dispersion of silica under a $g^*$ on the order of $10^{-3}$. More recently, we have demonstrated gravitational annealing in silica colloidal dispersion using centrifugation sedimentation.[29] After the formation of a crystal under 9G for 2 days, the centrifugation rotation velocity was increased to 50G and maintained for 5 days. We observed the disappearance of line-shaped defects. The corresponding $g^*$'s are on the order of $10^{-3}$ and $10^{-2}$, relatively small, respectively, for 9G and 50G. We wish to stress that a synergetic effect between the gravitational and patterned substrate effects works when using gravitational annealing/aging in colloidal epitaxy, leading to an enhanced reduction of defect.

## References and Notes

1  A. van Blaaderen, R. Ruel, P. Wiltzius, *Nature.* **1997**, *385*, 321.
2  J. Zhu, M. Li, R. Rogers, W. Mayer, R. H. Ottewill, STS-37 Space Shuttle Crew, W. B. Russel, P. M. Chaikin, *Nature* **1997**, *387*, 883.
3  A. Mori, Y. Suzuki, S.-i. Yanagiya, T. Sawada, K. Ito, *Molec.Phys.* **2007**, *105*, 1377; *Molec.Phys.* **2008**, *106*, 187.
4  A. Mori, *J. Cryst. Gwoth* **2011**, *318*, 66
5  A. Mori, Y. Suzuki, S. Matsuo, *World J. Eng.* **2012**, *9*, 37.
6  W. W. Wood, J. D. Jacobson, *J. Chem. Phys.* **1957**, *27*, 1207.
7  B. J. Alder, T. E. Temkin, *J. Chem. Phys.* **1957**, *27*, 1208.
8  W. G. Hoover, F. H. Ree, *J. Chem. Phys.* **1968**, *49*, 3609.
9  R. L. Davidchack, B. B. Laird, *J. Chem. Phys.* **1998**, *108*, 9452.
10  K. E. Davis, W. B. Russel, W. J. Glantsching, *Science* **1992**, *245*, 507.
11  K.-h. Lin, J. C. Crocker, V. Prasad, A. Schofield, D. A. Weitz, T. C. Lubensly, A. G. Yodh, *Phys. Rev. Lett.* **2000**, *85*, 1770.
12  D. K. Yi, E-M.. Seo, D.-Y. Kim, *Appl. Phys. Lett.*, **2003**, *80*, 255.
13  J. P. Hoogenboom, A. Yethiraj, A. K. van Langen-Suuring, J. Romijn, A. van Blaaderen, *Phys. Rev. Lett.* **2002**, *89*, 256104.
14  J. Zhang, A. Alsayed, K. H. Lin, S. Sanyl, F. Zhang, W. J. Pao, V. S. K. Balagurusmv, P. A. Heiney, A. G. Yodh, *Appl. Phys. Lett.*, **2002**, *81*, 3176.
15  W. K. Kegel, K. G. Dhont, *J. Chem. Phys.*, **2000**, *112*, 3431.
16  A. Mori, *World J. Eng.*, **2011**, *8*, 112.
17  in preparation for submission
18  The simulations were perfomed with double precision flating points. Therefore, there are eight significant digits (0.33842188 instead of 0.338). This note is applied to $L_x$=$L_y$ (25.092688) and $L_z$ (1000.0000). Note that for clarity ninth and tenth digits have been written for the diagonal length of the intersection of grooves.
19  A. Mori, S.-i. Yanagiya, Y. Suzuki, T. Sawada, K. Ito, *J. Chem. Phys.*, **2006**, *124*, 174507.
20  M. Megens, C. M. van Kats, P. Bösecke, W. L. Vos, *J. Appl. Cryst.*, **1997**, *30*, 637.
21  B. J. Ackerson, S. E. Pauling, . Johnson, W. van Megen, S. Undrwood, *Phys. Rev. E*, **1999**, *59*, 6903.
22  Y. Suzuki, T. Sawada, A. Mori, K. Tamura, *Kobuishi Ronbunshu* [in Japanese], **2007**, *64*, 161.
23  Y. Suzuki, T. Sawada, K. Tamura, *J. Cryst. Growth* **2011**, *318*, 780.
24  B. Cichicki, K. Hinsen, *Physica A* **1990**, *166*, 473.
25  E. Sanz, D. Marenduzzo, *J. Chem. Phys.* **2010**, *132*, 194102.
26  A. Mori, Y. Suzuki, *Molec. Phys.*, **2010**, *108*. 1731.
27  C. Dux, H. Versmold, *Phys. Rev. Lett.*, **1997**, *78*, 1811.
28  I. P. Dolbnya, A. V. Petukhov, D. G. A. L. Aarts, G. J. Vroege, H. N. W. Lekkerkerker, *Europhys. Lett.*, **2005**, *72*, 962.
29  Y. Suzuki, J. Endoh, A. Mori, T. Yabutani, K. Tamura, *Defect Diffusion Forum*, **2012**, *323-325*, 555.